\documentclass{aa}  
\usepackage{graphicx}
\usepackage{natbib}
\bibpunct{(}{)}{;}{a}{}{,} % to follow the A&A style
\usepackage{txfonts}
\begin{document}
   \title{Diffuse radio emission in the merging cluster \object{MACS~J0717.5+3745}: the discovery of the most powerful radio halo}
 \titlerunning{Diffuse radio emission in the cluster \object{MACS~J0717.5+3745}}
    % Three extremely steep spectrum radio relics
   %\subtitle{I. Overviewing the $\kappa$-mechanism}

   \author{R.~J. van Weeren\inst{1}
         \and H.~J.~A. R\"ottgering\inst{1}
          \and M. Br\"uggen \inst{2}
          \and A. Cohen \inst{3}
          }

   \institute{Leiden Observatory, Leiden University,
              P.O. Box 9513, NL-2300 RA Leiden\\
              \email{rvweeren@strw.leidenuniv.nl}
         \and             Jacobs University Bremen, PO Box 750 561, 28725 Bremen
         \and   Naval Research Laboratory, Code 7213, Washington, DC 20375, USA
             }

 %  \date{Received September 15, 2008; accepted March ??, 200?}

% \abstract{}{}{}{}{} 
% 5 {} token are mandatory
 
\abstract
   {Hierarchical models of structure formation predict that galaxy clusters grow via mergers of smaller clusters and galaxy groups, as well as through continuous accretion of gas. \object{MACS~J0717.5+3745} is an X-ray luminous and complex merging cluster, located at a redshift of $0.5548$. The cluster is suspected to host a bright radio relic, but up until now no detailed radio observations have been reported. Here we present Giant Metrewave Radio Telescope (GMRT) radio observations at 610~MHz of this cluster.
     }
   {The main aim of the observations is to study the diffuse radio emission within the galaxy cluster \object{MACS~J0717.5+3745} related to the ongoing merger.
   }
   {We have carried out GMRT $610$~MHz continuum observations of  \object{MACS~J0717.5+3745}. These are complemented by Very Large Array (VLA) archival observations at 1.4, 4.9 and 8.5~GHz. 
   }
  % results heading
  {We have discovered a radio halo in the cluster \object{MACS~J0717.5+3745} with a size of about 1.2~Mpc. The monochromatic radio power at 1400~MHz ($P_{1.4}$) is $5\times 10^{25}$~W~Hz$^{-1}$, which makes it the most powerful radio halo known till date. A 700~kpc radio structure, which we classify as a radio relic, is  located in between the merging substructures of the system. The global spectral index of radio emission within the cluster is found to be $-1.24 \pm 0.05$ between 4.9~GHz and 610~MHz. We derive a value of $5.8$~$\mu$G for the equipartition magnetic field strength at the location of the radio halo. The location of the relic roughly coincides with regions of the intra-cluster medium (ICM) that have a significant enhancement in temperature as shown by Chandra. The major axis of the relic is also roughly perpendicular to the merger axis. This shows that the relic might be the result of a merger-related shock wave, where particles are accelerated via the diffuse shock acceleration (DSA) mechanism. Alternatively, the relic might trace an accretion shock of a large-scale galaxy filament to the south-east. 
   }
   {}

   
   \maketitle

\section{Introduction}
According to hierarchical structure formation, galaxy clusters grow via mergers of smaller clusters and galaxy groups. During cluster mergers an enormous amount of energy is released. This energy can heat the intra-cluster medium and accelerate particles to highly relativistic energies \citep[e.g.,][]{2002ASSL..272....1S, 2008SSRv..134..311D, 2000ApJ...535..586T}. In the presence of magnetic fields these particles will emit diffuse synchrotron radiation, known as radio relics and halos \citep[see the reviews by][and references therein]{1996IAUS..175..333F, 2005AdSpR..36..729F,2008SSRv..134...93F}.
%\citep[see the review by][and references therein]{2008SSRv..134...93F}. 
Radio halos have a typical size of about 1 Mpc and the radio emission has a regular smooth morphology that follows the thermal X-ray emission from the ICM. 
Radio halos are unpolarized up to a few percent level. One of the main problems for understanding the presence of halos in clusters is the origin of the synchrotron-emitting electrons. The lifetime of these particles is much shorter than their diffusion time over the full extent of the halo \citep{1977ApJ...212....1J}. Two main possibilities have been proposed to explain the presence of the relativistic particles producing the radio halos: (i) the re-acceleration models, where relativistic particles are re-accelerated in situ by turbulence in the ICM generated by merger events \citep[e.g.,][]{2001MNRAS.320..365B, 2001ApJ...557..560P, 2003ApJ...584..190F}, and (ii) the secondary models, where the energetic electrons are secondary products of proton-proton collisions \citep[e.g.,][]{1980ApJ...239L..93D, 1999APh....12..169B, 2000A&A...362..151D}. Currently, the re-acceleration models provide a better explanation for the occurrence of radio halos since (i) halos seem to occur only in clusters which are undergoing a merger \citep[e.g., ][]{2007A&A...463..937V, 2008A&A...484..327V}, and (ii) the spectra of some radio halos show the expected high frequency cutoff \citep[e.g.,][]{2008Natur.455..944B}. This in contrast to secondary models that predict that halos should be present in all massive clusters and the spectral indices should follow simple power-laws, both of which appears to contradict observations.

Unlike halos, radio relics are irregular structures and can be highly polarized. Their sizes vary but relics with sizes of megaparsecs have been found in some clusters. These giant radio relics are thought to trace merger shocks, where particles are accelerated through the diffuse shock acceleration (DSA) mechanism \citep{1983RPPh...46..973D, 1998A&A...332..395E, 2001ApJ...562..233M}.

\label{sec:obs-reduction}
\begin{table*}
\begin{center}
\caption{Observations}
\begin{tabular}{lllll}
\hline
\hline
& GMRT 610 MHz & VLA 1.4 GHz & VLA 4.9~GHz & VLA 8.5~GHz \\
\hline
%Bandpass \& flux calibrator(s) & 3C 48, 3C 147 & 3C 48, 3C 147 & 3C 48, 3C 286 \\
%Phase calibrator(s)                   & 2225-049 &2225-049 & 0025-260 \\
Frequency (VLA band)                & 610~MHz &1385, 1465~MHz (L)&4835, 4885~MHz (C)& 8435, 8485~MHz (X)\\
Bandwidth		  & $2\times 16$~MHz& $2\times 50$~MHz & $2\times 50$~MHz & $2\times 50$~MHz \\
Channel width                      &125 kHz & 50~MHz&  50~MHz & 50~MHz\\
%Polarization			& RR& LL & RR+LL \\
%Sidebands			& USB+LSB & USB & USB \\
Observation dates				& 6, 8, Nov 2008& 15 Jul 2002, 22 Aug 1997  & 27 Nov 2001 & 22 Aug 1997  \\
Project Code & 15HRa01 & AE0147, AE0111&AH0748&AE0111 \\
%Integration time per visibility               	& 16.9 s &16.9 s& 16.9 s \\
Total on-source time		& $\sim 4$~hrs& 49~min  +18~min&  89~min & 8~min\\
VLA configuration   &&B+C&D&C\\
Beam size			&  $8.2\arcsec \times 6.0\arcsec$ & $6.1\arcsec \times 6.1.\arcsec$ & $17\arcsec \times 17\arcsec$ & $3.0\arcsec \times 2.9\arcsec$ \\
rms noise ($\sigma$)	& 78 $\mu$Jy beam$^{-1}$ & 49 $\mu$Jy beam$^{-1}$ & 22 $\mu$Jy beam$^{-1}$ & \ldots\\	%812 
Largest detectable angular scale &$\sim 4\arcmin$&15\arcmin&5\arcmin&$\sim1.5\arcmin$~$^{a}$\\
% 254 -->  297
\hline
\hline
\end{tabular}
\label{tab:observations}
\end{center}
$^{a}$ the largest detectable angular scale is about two times smaller than the theoretical value for short integration times\\
\end{table*}

An extreme example of a merging cluster is  \object{MACS~J0717.5+3745} \citep{2001ApJ...553..668E,2003MNRAS.339..913E}. 
This massive X-ray luminous cluster ($L_{X} = (24.6\pm0.3) \times 10^{44}$~erg~s$^{-1}$) is located at $z=0.5548$, with an overall ICM temperature of $11.6\pm0.5$~keV \citep{2007ApJ...661L..33E}.  The cluster shows pronounced substructure in optical images. A large-scale filamentary structure of galaxies connected with the cluster was discovered by  \cite{2004ApJ...609L..49E}.
Steep spectrum\footnote{$F_{\nu} \propto \nu^{\alpha}$, with $\alpha$ the spectral index} radio emission ($\alpha =-1.15$, between 1400 and 74~MHz) in the direction of the cluster is found in the NVSS \citep{1998AJ....115.1693C}, WENSS \citep{1997A&AS..124..259R} and VLSS \citep{2007AJ....134.1245C} surveys. The radio emission within the cluster was classified as a radio relic by \cite{2003MNRAS.339..913E} %, based on 5\arcsec resolution FIRST images. In addition, no obvious optical counterpart was found for this source. 
and no radio halo was found. %although this was mistakenly stated by \cite{2009ApJ...693L..56M}. 
\cite{2009ApJ...693L..56M} presented X-ray (Chandra) and optical observations (Hubble Space Telescope (HST), ACS; Keck-II, DEIMOS) of the cluster. Instead of just two clusters merging, as is the case for \object{1ES0657-56} \citep{2002ApJ...567L..27M}, they found the situation in \object{MACS~J0717.5+3745} to be more complex. A comparison of the galaxy and gas distribution with the radial velocity information showed the cluster to be undergoing a triple merger event. As pointed out by \citeauthor{2009ApJ...693L..56M}, the partial alignment of the merger axes for the subclusters suggests that these mergers are linked to the large-scale filament to the south-east of the cluster. ICM temperatures exceeding $20$~keV were found in some regions of the merging system. Near two of the merging subclusters decrements in the ICM temperature were observed which are probably remnants of cool cores.

One of the main questions is whether the radio relic in \object{MACS~J0717.5+3745} is the result of a large shock wave, where particles are accelerated via the DSA mechanism. According to  \citeauthor{2009ApJ...693L..56M}, the high temperature of the ICM in the cluster could to some extent be explained by shock heating from the ongoing mergers. However, the size of the high temperature ICM regions is very large ($\sim1$~Mpc), therefore they conclude that it is more likely we are seeing the result of contiguous accretion of gas along the cluster-filament interface.

In this paper we present 610~MHz continuum observations of \object{MACS~J0717.5+3745} with the GMRT. These observations are complemented by unpublished VLA archival observations. The layout of this paper is as follows. In Sect.~\ref{sec:obs-reduction} we give an overview of the observations and data reduction. The radio images, spectral index maps, and X-ray overlay are presented in Sec.~\ref{sec:results}. This is followed by a discussion and conclusions in Sects.~\ref{sec:discussion} and \ref{sec:conclusion}. 
Throughout this paper we assume a $\Lambda$CDM cosmology with $H_{0} = 71$ km s$^{-1}$ Mpc$^{-1}$, $\Omega_{m} = 0.3$, and $\Omega_{\Lambda} = 0.7$.

\section{Observations \& data reduction}
\object{MACS~J0717.5+3745} was observed with the GMRT as part of a larger sample of diffuse steep spectrum sources selected from the 74~MHz VLSS survey. Here we shortly describe the reduction of these observations, for further details the reader is referred to van Weeren et al. 2009 (submitted). The cluster was observed in the 610~MHz  band on 6 and 8 November, 2008. The total on-source time was about 4 hours, see Table~\ref{tab:observations}. The observations recorded both RR and LL polarizations in two intermediate frequency bands (IFs), each having 128~channels, with a total bandwidth of 32~MHz. %the upper (USB) and lower (LSB) sidebands (each 128~channels), with a total bandwidth of 32~MHz. 
We used the NRAO Astronomical Image Processing System (AIPS) for reducing the data. The data was flux and bandpass calibrated using the primary calibrators 3C48, 3C147 and 3C147. We used the \cite{perleyandtaylor} extension to the \cite{1977A&A....61...99B} scale for the absolute flux calibration. Special care was taken to remove radio frequency interference (RFI). To track gain variations during the observations the secondary calibrator 0713+438 was observed several times. After this initial calibration several rounds of phase selfcalibration were carried out followed by two final rounds of amplitude and phase selfcalibration. We used the polyhedron method \citep{1989ASPC....6..259P, 1992A&A...261..353C} for making images to minimize the effects of non-coplanar baselines. Both IFs were simultaneously gridded, imaged and cleaned. About two times the primary beam radius was covered with 199 facets to remove sidelobes from sources away from the field center.

\object{MACS~J0717.5+3745}  was observed with the VLA, in the L-band on 15 July 2002 in B-array, in C-band on 27 November 2001 in D-array, and in the L and X-bands on 22 August 1997 in C-array, see Table~\ref{tab:observations}. These observations were carried out in single channel continuum mode with two IFs each with a bandwidth of 50~MHz. The L-band B-array data was calibrated in the standard way using the flux calibrator 3C48 and phase calibrator 0713+438.  Several rounds of phase only selfcalibration were followed by a final round of amplitude and phase selfcalibration. The L and X-band C-array observations lacked a usable scan on a flux calibrator, so we could only use the secondary calibrator 0713+438. Flux calibration for the L-band data was carried out by an amplitude and phase selfcalibration against the B-array clean-component model, restricting the UV-range to avoid the inclusion of short baselines. This produced the desired result after checking the flux of the compact sources in the corrected images. The B and C-array L-band UV-data were then combined and imaged simultaneously. An extra round of amplitude and phase selfcalibration was carried out on the combined data. Flux calibration for the X-band data was not carried out as we only used the images to identify compact sources related to AGN. The calibrated C-band observations were taken from the NRAO VLA Archive Survey\footnote{This NVAS image was produced as part of the NRAO VLA Archive Survey, (c) AUI/NRAO. The NVAS can be browsed through http://www.aoc.nrao.edu/\~{}vlbacald/}. We carried out an additional round of phase selfcalibration. 

\section{Results: images, equipartition magnetic field \& spectral index maps}
\label{sec:results}

 % S21

\begin{figure}
    \begin{center}
      \includegraphics[angle = 90, trim =0cm 0cm 0cm 0cm,width=0.5\textwidth]{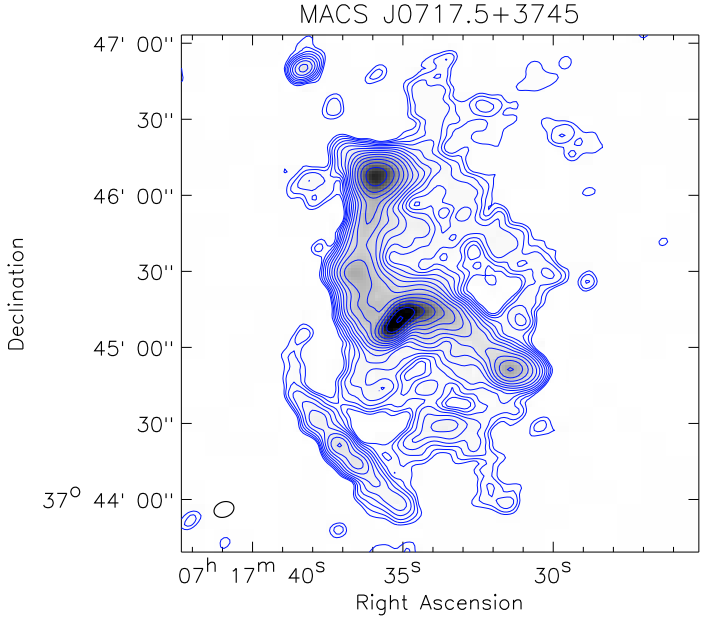}
       \end{center}
      \caption{GMRT 610~MHz radio image. Contour levels are drawn at $\sqrt{[1,2,4,6,8,16,32,...]}  \times 234$~$\mu$Jy~beam$^{-1}$. The image has a noise of 78~$\mu$Jy~beam$^{-1}$. The beam size is $8.2\arcsec \times 6.0\arcsec$ and is indicated in the left bottom corner.}
            \label{fig:radiomap21}
 \end{figure}

\begin{figure}
    \begin{center}
      \includegraphics[angle = 90, trim =0cm 0cm 0cm 0cm,width=0.5\textwidth]{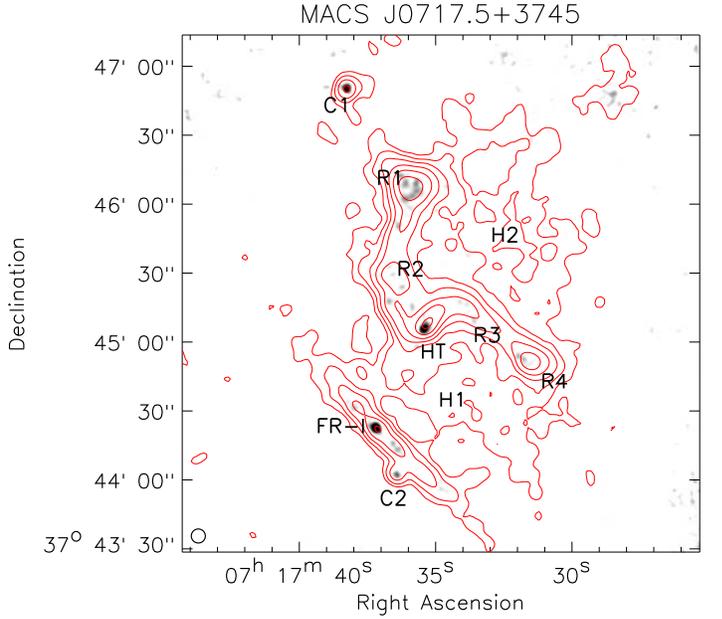}
       \end{center}
      \caption{Greyscale 8.5~GHz VLA C-array image with a beam of $3.0\arcsec\times2.9\arcsec$. Red contours are from the VLA B+C array 1.4~GHz radio image.   The image has a restoring beam size of $6.1\arcsec\times6.1\arcsec$ (shown at the bottom left corner) and and a rms noise of $49$~$\mu$Jy~beam$^{-1}$. Contours levels are drawn at $[1, 2, 4, 8, 16, 32,  ...]~\times 3\sigma_{\mathrm{rms}}$. Indicated in the figure are the radio halo (H1 \& H2), radio relic (R1$-$R4), central head-tail source (HT), source FR-I to the south-east, and two other compact sources (C1 \& C2).}
            \label{fig:radiomaplx}
 \end{figure}

%\begin{figure}
 %   \begin{center}
  %    \includegraphics[angle = 0, trim =0cm 0cm 0cm 0cm,width=0.49\textwidth]{temp+radio.png}
  %     \end{center}
  %    \caption{Temperature map of the cluster taken from \cite{2009ApJ...693L..56M}. GMRT 610~MHz contours in value are overlaid at levels of  $\sqrt{[1, 8, 32, 128, ...]}~\times 4\sigma_{\mathrm{rms}}$.}
   %         \label{fig: xray_s21}
% \end{figure}

%\begin{figure}
 %   \begin{center}
 %     \includegraphics[angle = 180, trim =0cm 0cm 0cm 0cm,width=0.49\textwidth]{xray_s21}
  %     \end{center}
  %    \caption{Chandra X-ray map overlaid with radio contours at 610~MHz from the GMRT. The color scale represents the X-ray emission from $0.5-7.0$~keV. The has been adaptively smoothed using the TARA$^{a}$ package using a minimal significance of $5$. Contour levels are drawn at  $\sqrt{[1, 8, 32, 128, ...]}$  $\times$ $0.312$~mJy~beam$^{-1}$.}
 %           \label{fig: xray_s21_old}
 %\end{figure}

%\begin{figure}
%    \begin{center}
%     \includegraphics[angle = 90, trim =0cm 0cm 0cm 0cm,width=0.5\textwidth]{xray_s21}
%       \end{center}
%      \caption{Chandra X-ray and HST image overlaid with radio contours at 610~MHz from the GMRT. The purple ($??-??$~keV) and blue ($??-??$~keV) colors represents the X-ray emission from Chandra. [insert Chandra bands] GMRT 610~MHz contours in value are overlaid in green at levels of  $\sqrt{[1, 8, 32, 128, ...]}~\times 4\sigma_{\mathrm{rms}}$.}
%           \label{fig: xray_s21_PR}
% \end{figure}

\begin{figure}
    \begin{center}
      \includegraphics[angle = 90, trim =0cm 0cm 0cm 0cm,width=0.5\textwidth]{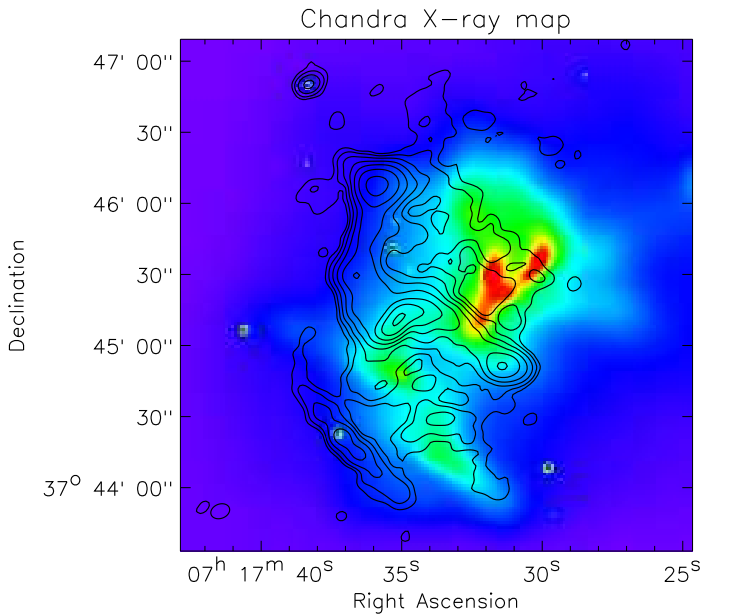}
       \end{center}
      \caption{Chandra X-ray map overlaid with radio contours at 610~MHz from the GMRT. The beam size is $8.2\arcsec \times 6.0\arcsec$ and is indicated in the left bottom corner. The color scale represents the X-ray emission from $0.5-7.0$~keV. The has been adaptively smoothed ($3\sigma$ confidence) using the \emph{ASMOOTH} algorithm \citep{2006MNRAS.368...65E}.    
     % TARA$^{a}$ package using a minimal significance of $5$. 
     Contour levels are drawn at $[1, 2, 4, 8, 16, 32,  ...]~\times$ $0.312$~mJy~beam$^{-1}$.}
%\footnote{http://www.astro.psu.edu/xray/docs TARA} 
%$^{a}$ http://www.astro.psu.edu/xray/docs/TARA
            \label{fig: xray_s21}
\end{figure}

Our GMRT radio image with a noise of 78~$\mu$Jy~beam$^{-1}$ is shown in Fig.~\ref{fig:radiomap21}. The 1.4, 4.9 and 8.5~GHz images are shown in Figs.~\ref{fig:radiomaplx} and \ref{fig:spix3} (the 4.9~GHz image as overlay on the spectral index map). The noise levels are 49 and 22~$\mu$Jy~beam$^{-1}$ for the 1.4 and 4.9~GHz images, respectively, see Table~\ref{tab:observations}. We have indicated the various sources found in Fig.~\ref{fig:radiomaplx}. 
The 610~MHz, 1.4 and 4.9~GHz images reveal a bright elongated source, defined as R, which we have further subdivided into R1 to R4. 
The source runs from north to south and then turns south-west at a compact source HT.  In the north and in the south-west, at the end of the elongated structure R, there are two regions with enhanced emission R1 and R4.  The linear size of the entire structure is about 700~kpc. On the basis of the spectral index map and location within the merging system we classify this source, except the central component (HT), as a radio relic (see Sects.~\ref{sec:spix}~and~\ref{sec:discussion}).  The bright, relatively compact region, HT, is located roughly halfway along this elongated source. To the southwest a straight elongated linear source (FR-I) is seen, most likely related to a nearby AGN. Around the radio relic (R), both to the south (H1) and north-west (H2), we detect faint diffuse radio emission with a total size of about 1.2~Mpc. %The described radio features above are also detected in the VLA 1.4 and 4.9~GHz images, see Fig.~\ref{fig:radiomaplx} and \ref{fig:spix3}. 

The 8.5~GHz image, Fig.~\ref{fig:radiomaplx}, shows four compact sources. HT is located on the elongated radio relic.  A hint of diffuse emission is seen from the northern brighter region R1. The second compact source is located halfway along the fainter linear structure (FR-I) to the south-east. Compact source C2 is located just south of FR-I, and compact source C1 is located to the north of R1. All these compact sources are related to AGN. No obvious optical counterpart was found for the central compact radio source (HT) by \cite{2003MNRAS.339..913E}. We identify an elliptical galaxy located within the cluster at RA~07$^\mathrm{h}$~17$^\mathrm{m}$~35$^\mathrm{s}$.5, DEC~+37\degr~45\arcmin~05\farcs5~as a likely counterpart. The source can be classified a ``head-tail'' source. The tail is not visible in the 8.5~GHz image due to its steep spectral index, but it is detected at all three frequencies below 8.5~GHz. The spectral index map clearly shows spectral steepening along the 20\arcsec~long tail as expected, see Fig.~\ref{fig:spix}. The south-east source FR-I with its compact core is associated with a bright elliptical foreground galaxy (RA~07$^\mathrm{h}$~17$^\mathrm{m}$~37$^\mathrm{s}$.2, DEC~+37\degr~44\arcmin~23\arcsec).  The faint linear extensions on both sides of the compact radio core probably make up a FR-I type radio source.%, although the morphology is somewhat peculiar as no lobes are visible.

\begin{figure}
    \begin{center}
      \includegraphics[angle = 90, trim =0cm 0cm 0cm 0cm,width=0.5\textwidth]{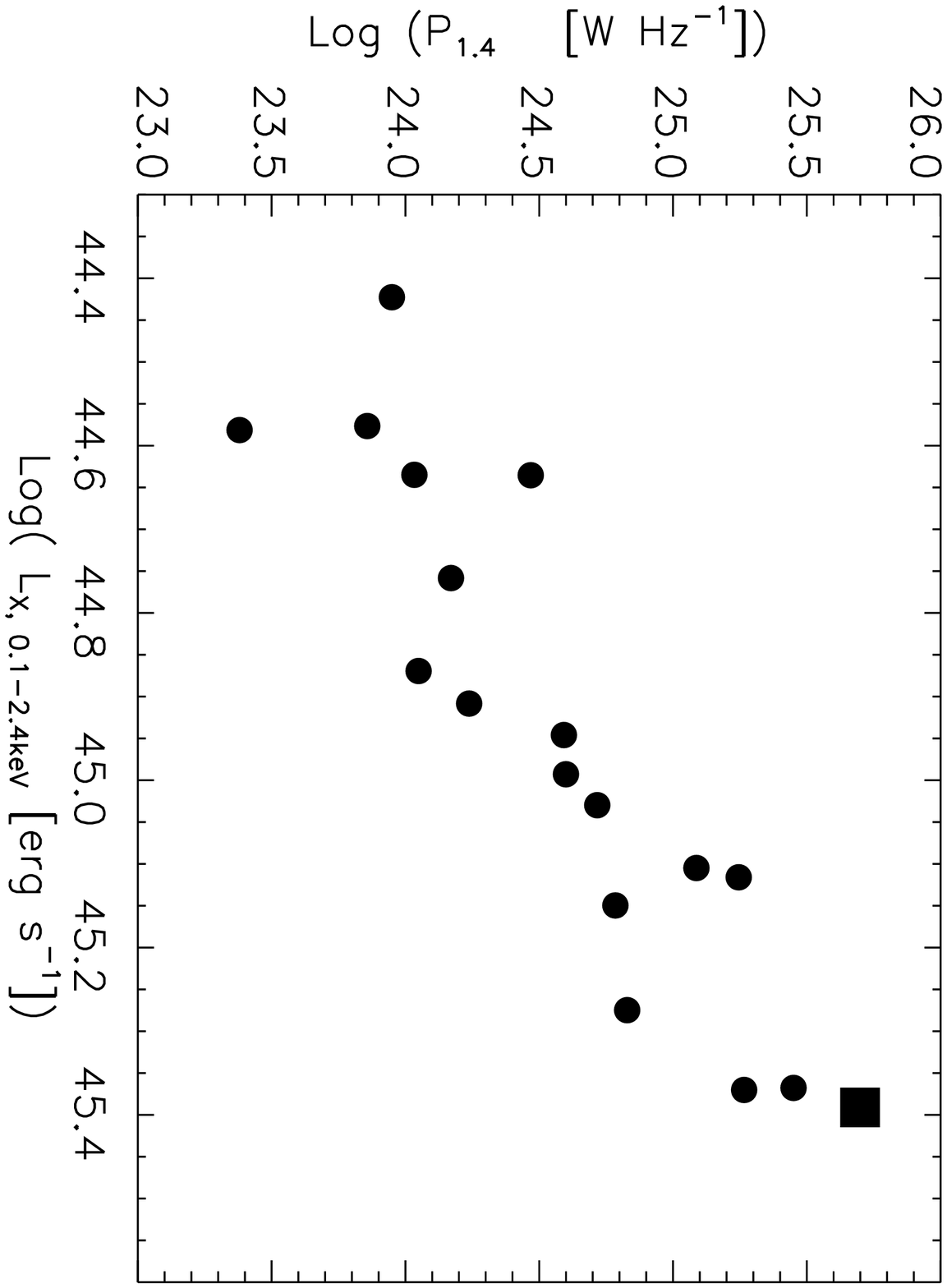}
       \end{center}
      \caption{X-ray luminosity versus 1.4~GHz radio power. The circles are from the halo list compiled by \cite{2006MNRAS.369.1577C}. The square represents the halo in \object{MACS~J0717.5+3745}.  } % 37 min exposure time
            \label{fig:L-P}
 \end{figure}

\begin{figure}
    \begin{center}
      \includegraphics[angle = 90, trim =0cm 0cm 0cm -0cm,width=0.5\textwidth]{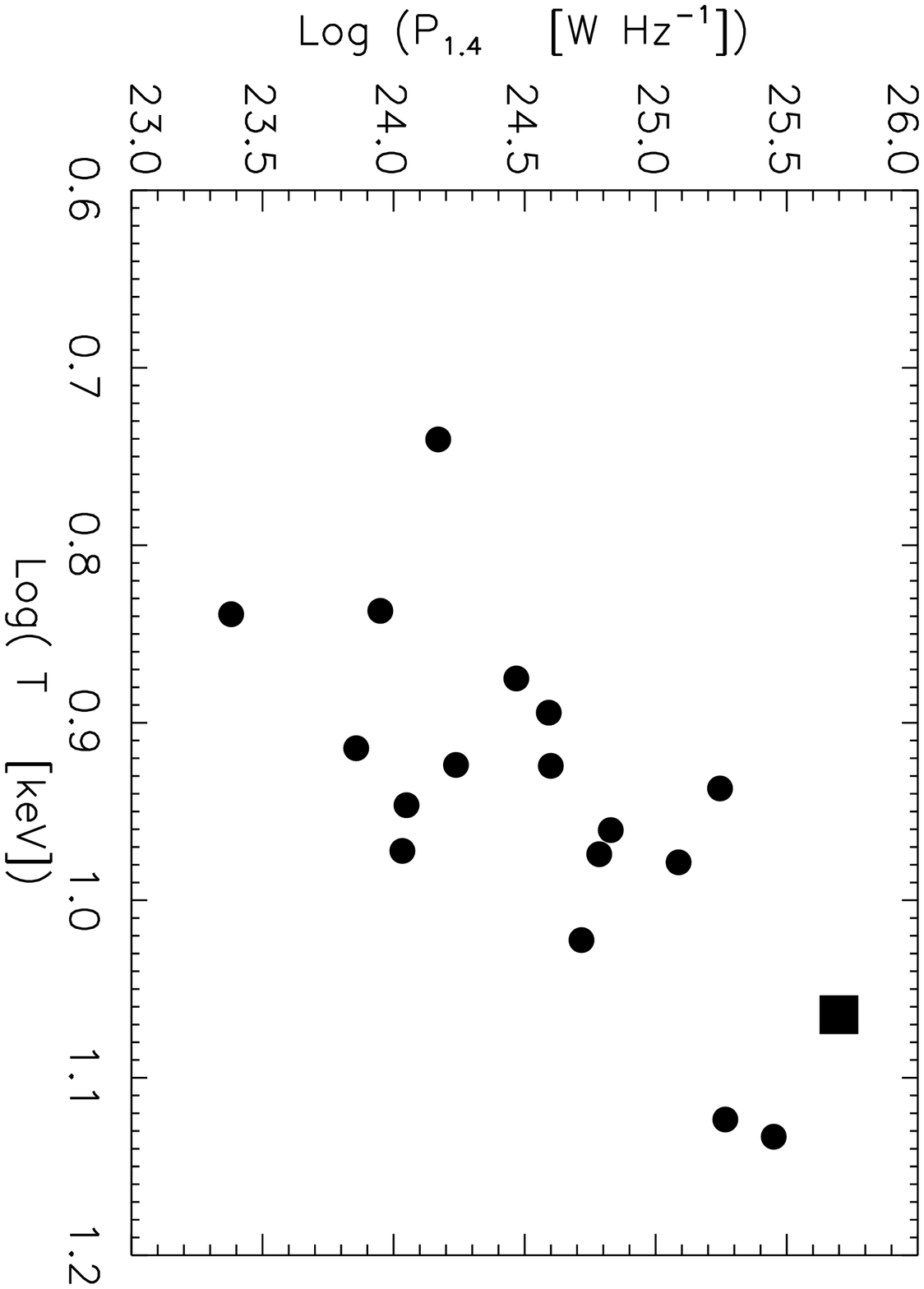}
       \end{center}
      \caption{ICM temperature versus 1.4~GHz radio power. The circles are from the halo list compiled by \cite{2006MNRAS.369.1577C}. The square represents the halo in \object{MACS~J0717.5+3745}.} % 37 min exposure time
            \label{fig:T-P}
 \end{figure}

The large-scale diffuse radio emission, H1+H2, is classified as a radio halo given the large size of about 1.2~Mpc and the emission following roughly the galaxy distribution and X-ray emission from the ICM. The monochromatic radio power ($P_{1.4}$) is estimated to be $(5 \pm 1) \times 10^{25}$~W~Hz$^{-1}$ using the 610~MHz image, a spectral index of $-1.1$ (see Fig.~\ref{fig:spix3}), an extent of 1.2~Mpc, and the AIPS task `TVSTAT'\footnote{TVSTAT measures the integrated flux in a user defined region of an image}. The uncertainty in the radio power is based on a spectral index error of about $0.3$. We have extrapolated the halo flux at the location of the bright radio relic using the average flux from the halo per unit surface area outside the relic region.  This increased the radio power by a factor of $1.23$. %Using the 1.4~GHz image we find a radio power of $6 \times 10^{25}$~W~Hz$^{-1}$, in agreement with the value derived from the 610~MHz image.  
This makes it the most powerful radio halo known as the radio power is higher than the halo in \object{1E0657-56} which has $P_{1.4}=3 \times 10^{25}$~W~Hz$^{-1}$ \citep{2000ApJ...544..686L}. In fact, we could have underestimated the radio power as we may have missed some additional diffuse emission due to our limited sensitivity on large angular scales. %This is especially true for the GMRT data were most short baselines had to be flagged because of RFI. 
The high radio power is in agreement with the X-ray luminosity-radio power ($L_{X}-P_{1.4}$) and temperature-radio power ($T-P_{1.4}$) correlations \citep{2000ApJ...544..686L,2002A&A...396...83E,2006MNRAS.369.1577C}, see Fig.~\ref{fig:L-P} and \ref{fig:T-P}.

An radio overlay on Chandra image is shown in Fig.~\ref{fig: xray_s21}. The radio relic is roughly located between the merging substructures as indicated by \cite{2009ApJ...693L..56M}. 

%The central radio structure coincides with regions having a significantly higher X-ray temperature $\gtrsim15$~keV and is located roughly between the subgroups B and C-D \citep[as labelled by][]{2009ApJ...693L..56M}. [expand/label depending on what image to show]


\subsection{Equipartition magnetic field strength}
The existence of a radio halo reveals the presence of magnetic fields within the cluster on a scale of about 1~Mpc. We can estimate the equipartition magnetic field strength on the basis of the minimum energy density of the magnetic field for the radio halo. We use the same procedure as in \cite{2009arXiv0908.0728V} and take $d=1.2$~Mpc (the depth of the source), $k=100$ (the ratio of energy in relativistic protons to that in electrons), and a spectral index of $-1.1$. Using these values and a flux of 6.8~$\mu$Jy~arcsec$^{-2}$ (integrated flux of the halo divided by the surface area) gives $B_{\mathrm{eq}}=3.2$~$\mu$G. For different values of $k$, the equipartition magnetic field strength scales with $(1+k)^{2/7}$. This method assumes fixed frequency cutoffs ($\nu_{\mathrm{min}}=10$~MHz and $\nu_{\mathrm{max}}=100$~GHz). This is not entirely correct, rather low and high energy cutoffs  ($\gamma_{\mathrm{min}},\gamma_{\mathrm{max}}$) for the particle distribution should be used \citep{1997A&A...325..898B, 2005AN....326..414B}. Assuming that $\gamma_{\mathrm{min}} \ll \gamma_{\mathrm{max}}$ and using a lower energy cutoff $\gamma_{\mathrm{min}} = 100$, the revised equipartition magnetic field strength $B^{\prime}_{\mathrm{eq}}$ is $5.8$~$\mu$G. For different values of $k$, the revised equipartition magnetic field strength ($B^{\prime}_{\mathrm{eq}}$) scales with $(1+k)^{1/(3-\alpha)}$. These equipartition magnetic field strength values are similar to those found for other radio halos \citep[e.g.,][]{1993ApJ...406..399G, 1999JKAS...32...75K, 2003A&A...397...53T, 2006AJ....131.2900C}. 
 
The equipartition energy density in the magnetic field can be compared to the thermal energy density for the cluster. For the density profile of the cluster we use the $\beta$-model \citep{1976A&A....49..137C} derived by \cite{2008ApJ...684..160M}.  We find that in the cluster center the thermal energy density is $1.9 \times 10^{-10}$~erg~cm$^{-3}$, the equipartition energy density in the magnetic field is $9.4 \times 10^{-13}$~erg~cm$^{-3}$. At a distance of 0.6~Mpc from the cluster center the thermal energy density drops to $5.9 \times 10^{-11}$~erg~cm$^{-3}$. The thermal energy density is about two orders of magnitude higher than the equipartition energy density in the magnetic field. This is consistent with the fact that a small fraction of the thermal energy is being used to accelerate particles to highly relativistic energies.

%The velocity of the shock wave can be estimated using the scaling relation given by \cite{1997MNRAS.286..257K}
%\begin{equation}
%V_{\mathrm{s}} = 1.75 \times 10^{3}  \mbox{ } \mathrm{ km s^{-1} } \mbox{ } \left( \frac{kT}{6.06  \mbox{ } \mathrm{ keV}}\right)^{1/2} \mbox{ .}
%\end{equation}
%Using the the average temperature ($T$) for the cluster of 11.6~keV gives $V_{\mathrm{s}}= 2.4 \times 10^{3}$~km~s$^{-1}$. Since the related

\subsection{Spectral index}
\label{sec:spix}
\begin{figure}
    \begin{center}
      \includegraphics[angle = 90, trim =0cm 0cm 0cm 0cm,width=0.5\textwidth]{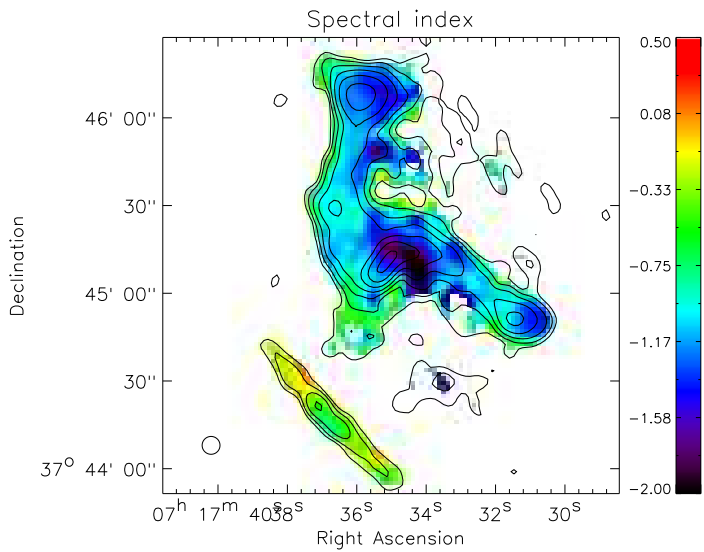}
       \end{center}
      \caption{Spectral index between 610~MHz and 1.4~GHz at a resolution of $6.1\arcsec\times6.1\arcsec$. Contour levels are from the 610~MHz image and drawn at levels of $[1, 2, 4, 8, 16, 32,  ...]~\times 0.468$~mJy~beam$^{-1}$. } % 37 min exposure time
            \label{fig:spix}
 \end{figure}
 \begin{figure}
    \begin{center}
      \includegraphics[angle = 90, trim =0cm 0cm 0cm 0cm,width=0.5\textwidth]{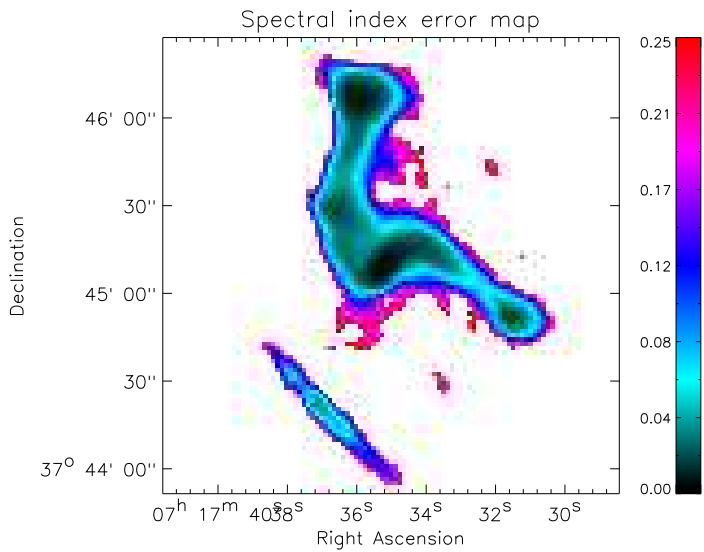}
       \end{center}
      \caption{Spectral index error map between 1.4~GHz and 610~MHz. The error map shows the uncertainties in the spectral index based on rms noise in both radio images. The absolute flux calibration uncertainty is not included.} 
            \label{fig:spix_error}
 \end{figure}

\begin{figure}
    \begin{center}
      \includegraphics[angle = 90, trim =0cm 0cm 0cm 0cm,width=0.5\textwidth]{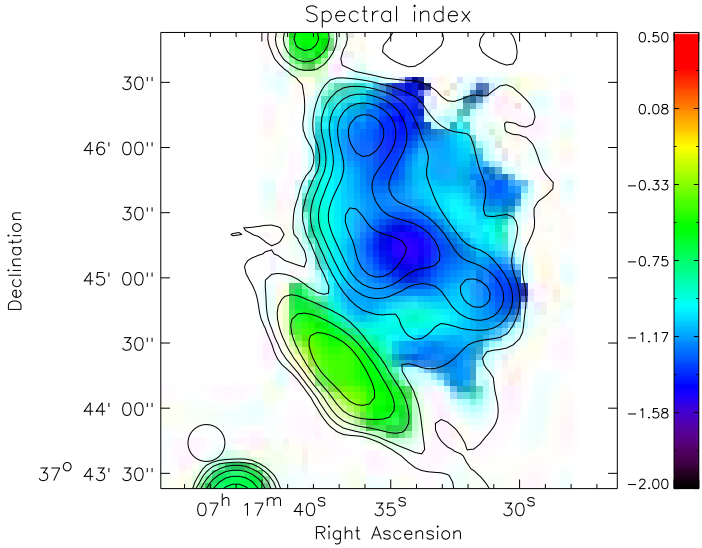}
       \end{center}
      \caption{Power-law spectral index fit between 0.61, 1.4, and 4.9~GHz at a resolution of $17\arcsec\times17\arcsec$ . Contours levels are from the 4.9~GHz image and drawn at levels of $[1, 2, 4, 8, 16, 32,  ...]~\times 0.14$~mJy~beam$^{-1}$}. 
            \label{fig:spix3}
 \end{figure}

\begin{figure}
    \begin{center}
      \includegraphics[angle = 90, trim =0cm 0cm 0cm 0cm,width=0.5\textwidth]{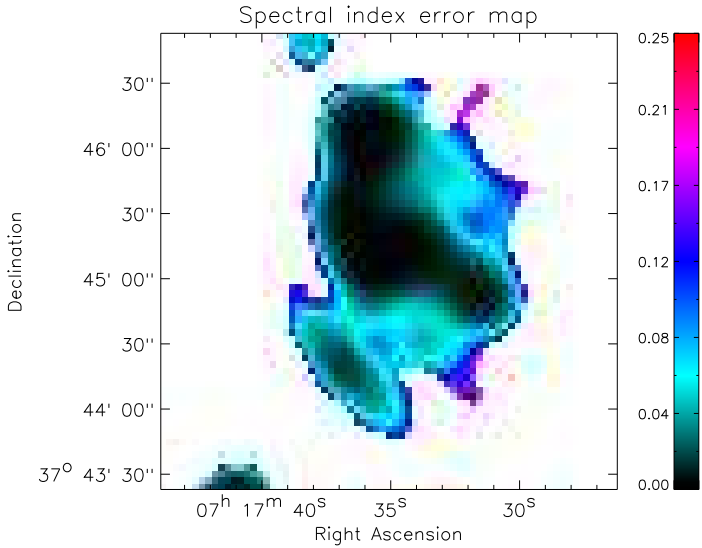}
       \end{center}
      \caption{Spectral index error map between 4.9~GHz, 1.4~GHz and 610~MHz. The error is the formal polynomial fitting error based on the rms noise in the three radio images. The absolute flux calibration uncertainty is not included}.
            \label{fig:spix3_error}
 \end{figure}

We have created a spectral index map between 1.4~GHz and 610~MHz at a resolution of 6.1\arcsec~shown in Fig.~\ref{fig:spix}. The uncertainties in the computed spectral index maps are shown in Fig.~\ref{fig:spix_error}. Pixels below $5\sigma_{\mathrm{rms}}$ in each of the images were blanked before computing the spectral index map. A second, lower resolution 17\arcsec~spectral index map, see Fig.~\ref{fig:spix3} and \ref{fig:spix3_error}, was created by fitting single power-law spectra to the combined images at 0.61, 1.4 and 4.9~GHz. The spectral index is only shown in regions where the flux is larger than $2.5\sigma_{\mathrm{rms}}$ at all three frequencies. Differences in the UV coverage and calibration errors may have resulted in additional errors in the spectral index.

%1.95e-1 1425
% 4.72e-1 610
% 3.62e-2 4860

Steepening along the tail of the central compact source (HT) is clearly visible. This is most likely caused by spectral ageing due to synchrotron and Inverse Compton (IC) losses. The spectral index of the head-tail source (HT) is steeper than that of the surrounding halo and relic. The origins of the electrons in HT and the halo/relic are very different. For HT the relativistic electrons are produced in an AGN, after that spectral ageing in the tail produces the steep spectrum. The electrons in the relic/halo are probably continuously being accelerated, the spectral index thus depends on the balance between ageing and acceleration. On top of that, the equipartition magnetic field strength in the tail is higher than that of the surrounding radio halo. 
The values of the (equipartition) magnetic field strength for the halo, tail, and equivalent magnetic field strength of the microwave background  \citep[$B_{\mathrm{IC}}= 3.25(1+z)^2$ in units of $\mu$G, e.g.,][]{1980ARA&A..18..165M, 2001AJ....122.1172S} are  3.2, 36, 7.8~$\mu$G, respectively. Comparing these values we find that the energy losses are higher in the tail compared to the halo, also resulting in a steeper spectral index.

In the high-resolution spectral index map (Fig.~\ref{fig:spix}), the spectral index in the two regions with enhanced emission (R1 and R4), at the north and south-west end of the relic, steepens to the west. This might be caused by spectral ageing behind a shock front, where particles are accelerated by the DSA mechanism, located on the west side of R1 and R4 (see also Sec.~\ref{sec:discussion_r}). This effect is visible to some extent at R2 in Fig.~\ref{fig:spix}. Alternatively, the relic traces a wider region with multiple shocks. Variations in the Mach number of the shocks would then cause variation in the (injection) spectral indices.  The overall spectrum is thus determined by the balance between spectral ageing, acceleration of particles, and the injection spectral index (determined by the Mach number of the shock). At present it is not possible to separate between these effects. %, but spectral ageing is most likely the dominant effect. 
Furthermore, spectral index variations on scales smaller than the beam size have been smoothed out. 

The low-resolution spectral index map is similar to the higher resolution one. However, the tail of the central compact radio source HT has a spectral index of only $-1.55$ compared to $-2.2$ in the high-resolution map. This is explained by the lower resolution of the second spectral index map. In this map the tail has been partly blended with the emission from regions surrounding it having a shallower spectrum.
The low-resolution spectral index map provides us with some indication of the spectral index of the radio halo. We find that on average $\alpha \approx -1.1$ for the radio halo. More observations are needed to better constrain this value and create a good spectral index map of the radio halo.

%The total integrated spectral index, excluding the south-east FR-I source, is $-1.04\pm 0.13$ between 610~MHz and 1.4~GHz, $-1.37\pm0.06$ between 1.4 and 4.9~GHz, and  $-1.24 \pm 0.05$ between 610~MHz and 4.9~GHz.  The errors are dominated by the absolute flux calibration uncertainty which we have taken to be 5\% for the VLA\footnote{VLA Calibrator Manual} and 10\% for the GMRT \citep{2004ApJ...612..974C}. 

 The total integrated spectral index, excluding the south-east FR-I source, is $-1.04\pm 0.13$ between 610~MHz and 1.4~GHz, $-1.37\pm0.06$ between 1.4 and 4.9~GHz, and  $-1.24 \pm 0.05$ between 610~MHz and 4.9~GHz.  The errors in the spectral indices are dominated by the absolute flux calibration uncertainty which we have taken to be 5\% for the VLA\footnote{VLA Calibrator Manual} and 10\% for the GMRT \citep{2004ApJ...612..974C}. These values for the spectral index are consistent with the spectral index of $-1.15 \pm 0.04$ derived from the 1.4~GHz NVSS, 325~MHz WENSS, and 74~MHz VLSS fluxes. The radio spectrum is thus relatively straight, although there is evidence for a small amount of spectral steepening at the higher frequencies due to spectral ageing. This indicates particle acceleration is currently still ongoing, as otherwise a strong high frequency cutoff would be expected.
  
The south-east linear structure (FR-I) has a relatively shallow spectral index of $-0.52\pm 0.13$ between 610~MHz and 4.9~GHz. In the high resolution map (Fig.~\ref{fig:spix}) the spectral index seems to flatten somewhat (from $-0.6$ to $-0.3$) away from the center of this source (the compact core seen in Fig.~\ref{fig:radiomaplx}). This could be explained by the presence of hotspots at the end of the lobes from the FR-I source. %Although the effect is not very significant as the error in the spectral index increases away from the center. 
The flattening is not clearly seen in the low-resolution spectral index map (Fig.~\ref{fig:spix3}). %It could be caused by the reduced sensitivity for emission on large spatial scales ($\gtrsim 1\arcmin$) in the high resolution 1.4~GHz image compared to the 610~MHz image. Alternatively, the spectral index is indeed flatter at both ends of the linear source. 


\section{Discussion}
 \label{sec:discussion}
\subsection{Alternative explanations for the elongated radio structures}
The bright elongated radio structure R could be a 700~kpc wide angle tail (WAT) source, with the enhanced regions of emission to the north and south-west (R1 and R4) the lobes/hotspots of this WAT. However, the spectral index map does not support such an explanation as there is no obvious connection with the central head-tail source (HT). The central head-tail source has a size of about 115~kpc,  much smaller than the total extent of the bright elongated radio structure.  Furthermore, the proposed lobes have a spectral index of about $-1.1$ and steepen towards the west. If the source was a 700~kpc WAT source, then a steepening of the spectral index along the proposed tails would be expected. This is not observed. 

Another possible interpretation for the fainter diffuse linear structure to the south-east (FR-I) is that of a second radio relic since the linear structure lies parallel to a ``ridge'' of X-ray emission visible in Chandra image (Fig.\ref{fig: xray_s21}). %Furthermore, the morphology is not like that of a typical FR-I source. 
The relic would then be seen in projection against the compact radio source located in front of it. However, we prefer a FR-I interpretation since (i) the source has a shallow spectral index of $-0.5$, and (ii) a central component is associated with a nearby active elliptical galaxy, also detected in the Chandra observations. This component lies halfway along the linear structure. High-resolution observations ($\lesssim 5\arcsec$) will be needed to confirm our proposed classification.

\subsection{Origin of the radio relic}
 \label{sec:discussion_r}
  \begin{figure}
    \begin{center}
      \includegraphics[angle = 90, trim =0cm 0cm 0cm 0cm,width=0.5\textwidth]{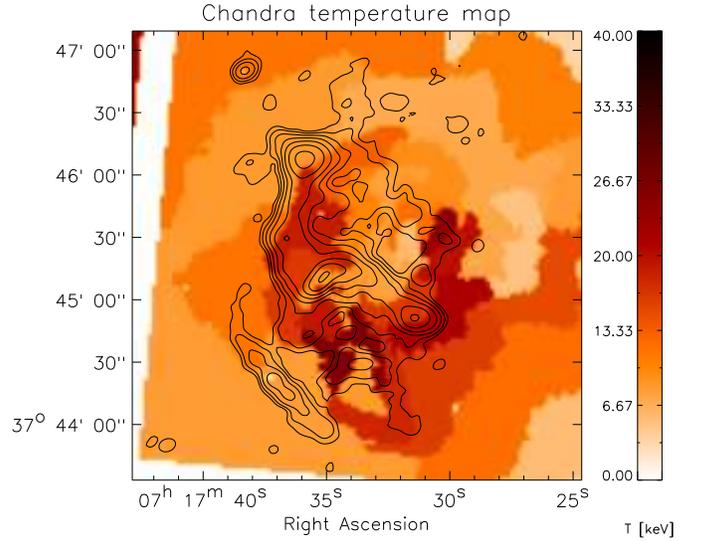}
       \end{center}
      \caption{Temperature map \citep[Fig.~2 from][]{2009ApJ...693L..56M} overlaid with 610~MHz radio contours. Contour levels are drawn at $[1, 2, 4, 8, 16, 32,  ...]~\times$ $0.312$~mJy~beam$^{-1}$. The beam size is $8.2\arcsec \times 6.0\arcsec$ and is indicated in the left bottom corner. }
            \label{fig:Xtemp}
 \end{figure}
 
The projected merger axes of the subgroups in the plane of the sky run in a direction that all point towards the large-scale galaxy filament to the south-east. The large elongated radio source (R) has an orientation roughly perpendicular to the merger axis which is expected for a relic tracing a large shock front, where electrons are accelerated via the DSA mechanism in a first-order Fermi process \citep[e.g.,][]{2008MNRAS.391.1511H}. Since the relic is expected to trace a shock wave in the ICM, it is interesting to compare our radio image with the low-resolution temperature maps that have been published by \cite{2009ApJ...693L..56M}, see Fig.~\ref{fig:Xtemp}. The south part of the relic (R3) is located north, but relatively close to a region %B1 \citep[as defined in Fig. 2 from][]{2009ApJ...693L..56M}  
having $kT \sim 25$~keV. The northern part of the relic (R1 and R2) directly follows a %A12, A6 and A9 
$kT \sim 20$~keV extension to the north of cluster. The spectral index for the north part of the relic seems to steepen away from these hot ICM regions. This is expected in the case of an outwards (to the east) traveling shock front where particles lose energy by synchrotron and IC losses in the post-shock region. The southern part of the relic (R4) also passes through regions %A8, A3, and A10 
having $kT \sim 20$ keV. Close to HT the relic does not directly trace any regions with a high temperature. However, the temperature map from \citeauthor{2009ApJ...693L..56M} has a fairly low resolution so a one-to-one comparison with the position of the relic is currently not possible. The temperature map also does not have the resolution to identify the location of a shock wave related to the ongoing merger.  

 Because of the large 1~Mpc extent of the high-temperature regions, \citeauthor{2009ApJ...693L..56M}  conclude that it is more likely that the ICM is heated as a result of contiguous accretion of gas along the cluster-filament interface. In this case, the radio relic might trace an accretion shock rather than a merger shock. Merger and accretion shocks generally differ in their location with respect to the cluster and also in Mach numbers \citep{2000ApJ...542..608M, 2002MNRAS.337..199M, 2003ApJ...593..599R, 2006MNRAS.367..113P, 2007MNRAS.375...77H, 2008MNRAS.391.1511H}. Accretion shocks have high mach numbers ($\mathcal{M} \gg 1$) and process the low-density, un-shocked inter-galactic medium (IGM). Merger shocks have lower Mach numbers and occur within the cluster, at radii smaller than about 1~Mpc. The relic in \object{MACS J0717.5+3745} is located relatively close to the cluster center, although this may (partly) be the result of projection. The spectral index over the relic varies roughly between -0.75 and -1.5. Therefore the radio observations do not exclude any of the two scenarios. Additional Chandra observations will be needed to establish the location of the proposed shock front and to link this to the location of the radio relic.

% -------------------- OTHER EXPLANATIONS

% FUTURE POL OBS
Polarization observations of the radio relic will be important to investigate the relation between the relic and the suggested shock wave that has either been created by the merger or by the accretion of gas. Shocks waves in the ICM/IGM can compress and order magnetic fields. If the magnetic field is seen under some angle (between the normal of the shock front and the line of sight) the projected magnetic field is expected to be parallel to the major axis of the relic \citep{1998A&A...332..395E}. The polarization E-vectors are then perpendicular to the shock front. This is indeed what has been observed for the double relics in \object{Abell~1240} and \object{Abell 2345} by \cite{2009A&A...494..429B} for example. Therefore future polarization observations of the relic in  \object{MACS~J0717.5+3745} will be important to test the shock wave scenario, where particles are accelerated by the DSA mechanism (either created by a cluster merger or by accretion of gas).

\section{ Conclusions}

\label{sec:conclusion}

We have presented GMRT 610~MHz radio observations of the complex merging cluster \object{MACS~J0717.5+3745}. We confirm the presence of a bright peripheral radio relic. We also report on the presence of additional diffuse emission in the cluster in the form of a radio halo, with $\alpha \approx -1.1$. The radio halo has the highest radio power ($P_{1.4}$) known to date. This radio power is in line with scaling relations, relating it to X-ray luminosity and ICM temperature. The integrated spectral index of the radio emission within the cluster is $-1.24 \pm 0.05$ between 610~MHz and 4.9~GHz. We have derived an equipartition magnetic field strength of $3.2$~$\mu$G for the radio halo. A somewhat higher value of 5.8~$\mu$G is obtained by using lower and higher energy cutoffs instead of fixed frequency cutoffs.

The location of the bright peripheral relic roughly coincides  with regions in the cluster having higher temperatures. The major axis of the relic is orientated approximately perpendicular to the merger axis of the system. This shows that the relic is probably the result of a large-scale shock wave within the cluster were particles are accelerated via the DSA mechanism. %The global spectral index of the radio emission within the cluster is $-1.04\pm 0.13$ between 610~MHz and 1.4~GHz and $-1.37\pm0.06$. 
Spectral index maps created using additional archival VLA observations show the presence of a 115~kpc head-tail source located roughly halfway the bright radio relic. 

%The spectral index seems to steepen somewhat to the west for the radio relic. This is expected if the relics traces an outwards traveling shock front related to the mergers taking place.
Follow-up polarization observations will be important to test the prosed shock wave scenario for the origin of the relic. High sensitivity radio observations at 8~GHz with a resolution $\lesssim 2\arcsec$ will allow a more detailed morphological study of the radio relic. This would also provide spectral indices with lower errors due to the larger spectral baseline and allow a search for spectral breaks or curvature.

Deep Chandra observations will be needed to create high spatial resolution temperature maps and pinpoint the location of the proposed shock wave. 
These observations might also be able to discriminate between the merger and accretion shock scenarios.

%Furthermore detailed temperature maps will be important to investigate the link between particle acceleration sites and the occurrence of diffuse radio emission in the cluster.

\begin{acknowledgements}
We would like to thank the anonymous referee for useful comments. We thank the staff of the GMRT who have made these observations   possible. GMRT is run by the National Centre for Radio Astrophysics of the Tata Institute of Fundamental Research.
The National Radio Astronomy Observatory is a facility of the National Science Foundation operated under cooperative agreement by Associated Universities, Inc. We would like to thank H.~Ebeling for providing the Chandra X-ray image and temperature map.

\end{acknowledgements}

\bibliographystyle{aa}
\bibliography{ref_zwcl}
\end{document}